\documentstyle[preprint,aps,floats,epsfig]{revtex}
\tightenlines
\begin{document}
\draft

\preprint{\vbox{\hbox{JLAB-THY-00-31}}}

\title{ Unfactorized versus factorized calculations for
$^2 H (e,e'p)$ reactions at GeV energies}

\author{Sabine Jeschonnek}

\address{\small \sl 
Jefferson Lab, 12000 Jefferson Avenue, Newport News, VA 23606 }

\date{\today}
\maketitle

\begin{abstract}
In the literature, one often finds calculations of $(e,e'p)$ reactions
at GeV energies using the factorization approach.  Factorization
implies that the differential cross section can be written as the
product of an off-shell electron-proton cross section and a distorted
missing momentum distribution. While this factorization appears in the
non-relativistic plane wave impulse approximation, it is broken in a
more realistic approach. The main source of factorization breaking are
final state interactions. In this paper, sources of factorization
breaking are identified and their numerical relevance is examined in
the reaction $^2H(e,e'p)$ for various kinematic settings in the GeV
regime.  The results imply that factorization should not be used for
precision calculations, especially as unfactorized calculations are
available.

\end{abstract}

\section{Introduction}

The study of electron scattering from nuclei has brought us many
insights over the past decades, starting with Hofstaedter's classic
inclusive electron scattering experiments which determined charge
radii, and continuing to the modern day coincidence experiments which
yield detailed information on the nuclear responses which allow us to
study the short range structure of nuclei and the properties of
nucleons in the nuclear medium.

In the past years, with the advent of high duty cycle machines with
several GeV of beam energy, coincidence experiments with GeV energy
and three-momentum transfers have become feasible and are carried out
mainly at Jefferson Lab, and with some limitations in beam energy also
at MAMI and Bates.  These high energy and momentum transfers permit us
to study the transition from hadronic degrees of freedom to
quark-gluon - or quark and flux tube - degrees of freedom in the
nucleus. Naturally, the interpretation of the data and the extraction
of the desired information is feasible only with a detailed knowledge
of the whole reaction. The general philosophy is that if we can not
describe a data set with the best ``conventional nuclear physics''
calculation, which would involve just hadronic degrees of freedom -
one-body currents and meson exchange currents, isobars, initial and
final state correlations - we would see evidence for genuine quark
effects in the nucleus.  The main practical problem for the time being
is that for the realm of several GeV, where the chance to see quark
effects is expected to be highest, the ``conventional nuclear
physics'' calculations have not yet been fully developed.

The main problems are a consistent or at least realistic description
of the final and initial hadronic states, proper inclusion of
relativistic effects \cite{origins}, especially the development of
relativistic meson exchange currents \cite{billmec}, and isobar
states. While all this has been achieved and worked out in great
detail over the past twenty years for the regime of lower energy and
three-momentum transfers of the order of a few hundred MeV, see e.g.
\cite{raskintwd,boffi,arenhoevel,janlowen}, a lot of work still needs to
be done in the GeV regime. In this regime, one needs new techniques:
the nature of the $NN$ interaction changes and takes on a
diffractive character, a description in terms of partial waves becomes
impractical, particle production is possible and indeed the most
frequent process, and relativity plays an important role.

Currently, even in the best available theoretical calculations,
approximations are necessary. However, in many cases, even more
approximations than necessary are used, and one of them, the
approximation of {\em factorization}, is the topic of this paper.
Numerical results for the validity of this approximation in $(e,e'p)$
reactions at GeV energies presented in this paper are for deuteron
targets and have been obtained using the Argonne V18 wave function
\cite{AV18}. Note that the factorization approximation is in general
not used in calculations at lower energies, see
e.g. \cite{raskintwd,boffi,arenhoevel,janlowen}.

This paper is organized as follows: after giving a brief overview over
the general formalism and notation in section \ref{subsecover}, I will
discuss the factorization approximation in section \ref{secwhatisfac},
and illustrate the mechanism of factorization breaking by final state
interaction with the simple example of a strictly non-relativistic
one-body current. In section \ref{secres}, I present numerical
examples for factorization breaking with a relativistic current
operator, and then summarize my results in the last section.

\subsection{Brief overview over the formalism and notation}
\label{subsecover}

In order to compare the full calculation with the factorized approach,
I start by introducing some notation and giving a brief summary of
the basic formalism of $(e,e'p)$ reactions. More details can be found
in \cite{raskintwd,dmtrgross}.

The differential cross section in the lab frame is 

\begin{eqnarray}
\left ( \frac{ d \sigma^5}{d \epsilon' d \Omega_e d \Omega_N}
\right ) _{fi}^h  & = & 
\frac{m_N \, m_f \, p_N}{8 \pi^3 \, m_i} \, \sigma_{Mott} \, 
f_{rec}^{-1} \, \nonumber \\
& & \Big[  v_L R_{fi}^L +   v_T R^T_{fi}
 + v_{TT} R_{fi}^{TT} + v_{TL} R_{fi}^{TL} 
  \nonumber \\
& & +  h \left ( v_{T'} R_{fi}^{T'} +  v_{TL'} R_{fi}^{TL'}
\right ) \Big] \, ,
\label{wqdef}
\end{eqnarray}
where $m_i$, $m_N$ and $m_f$ are the masses of the target nucleus, the
ejectile nucleon and the residual system, $p_N$ and $\Omega_N$ are the
momentum and solid angle of the ejectile, $\epsilon'$ is the energy of
the detected electron and $\Omega_e$ is its solid angle.  The helicity
of the electron is denoted by $h$.  
The Mott cross section is 
\begin{equation}
\sigma_{Mott} = \left ( \frac{ \alpha \cos(\theta_e/2)}
{2 \varepsilon \sin ^2(\theta_e/2)} \right )^2 \,,
\end{equation}
and the recoil factor is given by
\begin{equation}
f_{rec} = | 1+ \frac{\omega p_x - E_x q \cos \theta_x}
{m_i \, p_x} | \, .
\end{equation}
The coefficients $v_K$ are the
leptonic coefficients, and the $R_K$ are the response functions
which are defined by

\begin{eqnarray}
R_{fi}^L & \equiv & | \rho (\vec q)_{fi}|^2 \nonumber \\
R_{fi}^T & \equiv & | J_+ (\vec q)_{fi}|^2 
+ | J_- (\vec q)_{fi}|^2  \nonumber  \\
R_{fi}^{TT} & \equiv &  2 \, \Re \, \big[ J_+^* (\vec q)_{fi} \,
J_- (\vec q)_{fi} \big] \nonumber  \\
R_{fi}^{TL} & \equiv & - 2 \, \Re \, \big[ \rho^* (\vec q)_{fi} \,
( J_+ (\vec q)_{fi} - J_- (\vec q)_{fi}) \big] \nonumber \\
R_{fi}^{T'} & \equiv & | J_+ (\vec q)_{fi}|^2 -
 | J_- (\vec q)_{fi}|^2  \nonumber  \\
R_{fi}^{TL'} & \equiv & - 2 \, \Re \, \big[ \rho^* (\vec q)_{fi} \, 
( J_+ (\vec q)_{fi} + J_- (\vec q)_{fi}) \big] \, , 
\label{defresp}
\end{eqnarray}
where the $J_{\pm}$ are the spherical components of the
electromagnetic current.  For my calculations, I have chosen the
following kinematic conditions: the z-axis is parallel to $\vec q$,
the missing momentum is defined as $\vec p_m \equiv \vec q - \vec
p_N$, so that in Plane Wave Impulse Approximation (PWIA), the missing
momentum is equal to the negative initial momentum of the struck
nucleon in the nucleus, $\vec p_m = -\vec p$. I denote the angle
between $\vec p_m$ and $\vec q$ by $\theta_m$, and the term ``parallel
kinematics'' indicates $\theta_m = 0^o$, ``perpendicular kinematics''
indicates $\theta_m = 90^o$, and ``anti-parallel kinematics'' indicates
$\theta_m = 180^o$.  Note that both this definition of the missing
momentum and the definition with the other sign are used in the
literature.  In this paper, I assume that the experimental conditions
are such that either the kinetic energy of the outgoing nucleon and
the angles of the missing momentum, $\theta_m$, and the azimuthal angle
$\phi_m$, are fixed, or that the transferred energy $\omega$, the
transferred momentum $\vec q$, and the azimuthal angle $\phi_m$, are
fixed. In the former case, the transferred energy and momentum change
for changing missing momentum, in the latter situation, the kinetic
energy and polar angle of the outgoing proton change for changing
missing momentum.

\section{What is factorization?}
\label{secwhatisfac}

Factorization appears naturally in the non-relativistic plane wave
impulse approximation (PWIA) (see e.g. \cite{cdp}). There, one can
describe the differential cross section for the full process as
proportional to the product of the electron-proton cross section and
the spectral function. The spectral function $S (E, \vec p)$ describes
the probability to find a proton with a certain energy $E$ and
momentum $\vec p$ inside the nucleus.
\begin{equation}
\frac{ d^6 \sigma}{d \epsilon' d \Omega_e d \Omega_N d E_N} =
\frac{m_N \, m_f \, p_N}{E_f} \, \sigma_{eN} \, 
\, S (E, \vec p) \,.
\end{equation}
After integrating over the ejected nucleon's energy one finds
\begin{equation}
\frac{ d^5 \sigma}{d \epsilon' d \Omega_e d \Omega_N} = 
\frac{m_N \, m_f \, p_N}{m_i} \, \sigma_{eN} \, 
f_{rec}^{-1} \, n (\vec p) \,,
\label{pwiafactor}
\end{equation}
where $n(\vec p)$ is the momentum distribution. The
$eN$ cross section is given by
\begin{equation}
\sigma_{eN} = \sigma_{Mott} \sum_K v_K R_K^{single \, nucleon} \,,
\end{equation}
and the single nucleon responses are related to the nuclear responses
by
\begin{equation}
R_K^{nucleus} = (2 \pi )^3 \, R_K^{single \, nucleon} \, n(\vec p)
\end{equation}
so that one has in total:
\begin{equation}
\frac{ d^5 \sigma}{d \epsilon' d \Omega_e d \Omega_N} = 
\frac{m_N \, m_f \, p_N}{m_i} \,
f_{rec}^{-1} \,  \sigma_{Mott} \, n (\vec p) \,
\sum_K v_K R_K^{single \, nucleon} \,.
\end{equation}

These simple and intuitive results are valid only under the special
conditions of the non-relativistic PWIA: 1) There is no final state
interaction between the ejected nucleon and the residual nucleus. 2)
The negative energy states present in a relativistic treatment are
neglected. 3) The nucleon struck by the virtual photon is the one which
is detected in coincidence with the electron. The last
condition is commonly referred to as
impulse approximation (IA).

The main culprit for breaking factorization is the final state
interaction, which is always present in the general case. The
factorization breaking introduced by relaxing the other two conditions
are a bit more subtle. The size of the factorization breaking
introduced by negative energy states and terms beyond the impulse
approximation depends on the observable and kinematic region one
considers. For the unpolarized cross section in the GeV region, these
effects and the associated factorization breaking are small.  The
negative energy states which are present in the relativistic treatment
lead to a breaking of factorization, as was pointed out in
\cite{wallynf,rpwia,gardner}. An illustrative example for the case of
a deuteron target is shown in \cite{origins}. There, it was also shown
that a relativistic, positive-energy current operator reproduces the
fully relativistic, manifestly covariant result for missing momenta up
to 400 MeV/$c$, and that deviations for higher missing momenta stem
from off-shell effects and not from the negative energy states.

The assumption of the impulse approximation is quite good for high
energy and momentum transfers. The additional graph present in the
Born approximation (BA) describes the situation that the nucleus breaks up
and a nucleon that did not interact with the virtual photon is
detected. When high energies and momenta are transferred, it is very
unlikely that the initially struck nucleon transfers all of its
momentum to another nucleon in the final state interaction, or that
another nucleon could have such high momentum already in the ground
state. Therefore, in the region of GeV energy and momentum transfers
relevant to this paper, a full Born approximation calculation differs from the
impulse approximation calculation at most by a few percent.

As stated above, the final state interactions are the main source of
factorization breaking in the kinematics considered in this paper.
Nevertheless, one finds many calculations in the GeV regime assuming
factorization, even in the presence of final state interactions
\cite{fs,kolya,ciofi,myoldpapers}:
\begin{equation}
\frac{ d^5 \sigma^{factorized}}{d \epsilon' d \Omega_e d \Omega_N} = 
\frac{m_N \, m_f \, p_N}{m_i} \, \sigma_{eN} \, 
f_{rec}^{-1} \, n^{distorted} (\vec p, \vec p_m) \,,
\end{equation}
where the distorted missing momentum distribution is given by
\begin{equation}
n^{distorted} (\vec p, \vec p_m) = 
\frac{1}{(2 \pi)^3} \, \bar {\sum_f} |{\cal{M}}_f|^2
\end{equation}
with
\begin{eqnarray}
{\cal{M}}_f &= &<f|\hat S_{FSI}|i> \nonumber \\
&=& \int d \vec R_1 \dots d \vec R_{A-1} \,
\Psi^*_f ( \vec R_1 \dots\vec R_{A-2}) \,
\hat S_{FSI} (\vec r_1, \dots, \vec r_A) \,
\exp (i \vec p_m \vec R_{A-1}) \,
\Psi_i ( \vec R_1 \dots\vec R_{A-1}) \,.
\end{eqnarray}
Here, $\hat S_{FSI}$ is the final state interaction operator. Jacobi
coordinates are denoted by $\vec R$, the laboratory system coordinates
are denoted by $\vec r$.  The factorization approximation reduces the
numerical effort as only one integral needs to be evaluated.
In the unfactorized approach, 
 every part of the electromagnetic
current operator is evaluated separately, and the cross section is
built up from the different response functions based on the matrix
elements $<f|\hat S_{FSI} \, J_{em}|i>$, as written out in
Eq. (\ref{defresp}). Of course, when assuming factorization, any
difference in the behavior of the different response functions is
neglected. There are some cases when this obviously cannot work,
e.g. for the fifth response, $R_{TL'}$, which is measurable only with
a polarized electron beam. In the absence of final state interaction,
the fifth response is identically zero. While it is quite clear from this
example that factorization does not work for polarization observables,
the quality of the factorization approximation for the unpolarized
cross section and response functions is not clear a priori, and is
investigated in this paper. It largely depends on which components of
the current operator are involved in calculating a specific
observable.

\subsection{A simple example of factorization breaking}

\begin{figure}[!h,t]
\begin{center}
\leavevmode
\epsfig{file = 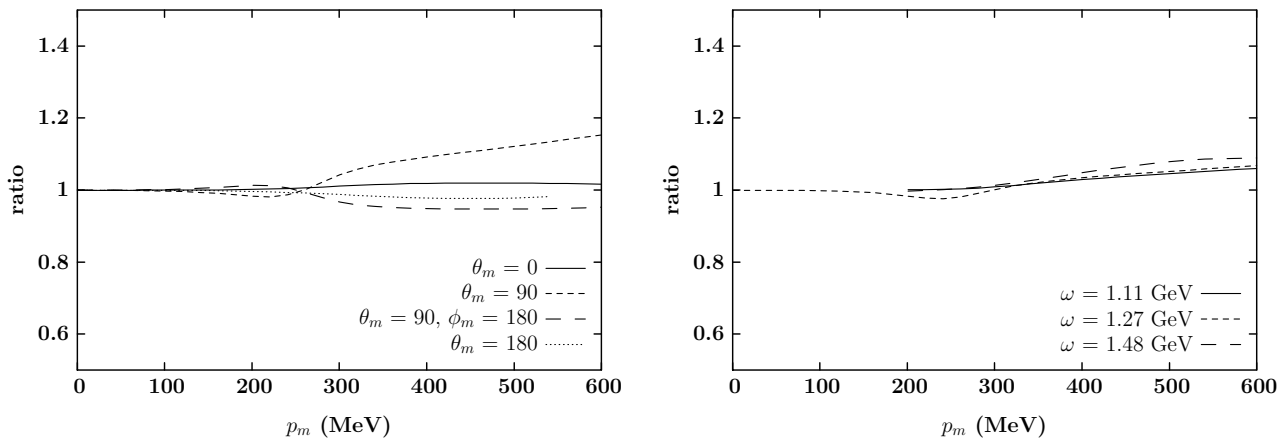, height = 16cm}
\end{center}
\vspace{-9cm}
\caption{The ratio of the cross section calculated in the
factorization approximation to the unfactorized cross section for
different kinematic settings.  The non-relativistic form of the
current operator was employed. The left panel shows the
ratio for fixed kinetic energy of 1 GeV of the outgoing proton
and various fixed angles of the missing momentum, the
right panel shows the ratio for fixed 3-momentum transfer $q = 2.0$
GeV/c and different values of the fixed energy transfers $\omega$:
1.11 GeV, 1.27 GeV, and 1.48 GeV.}
\label{figratnrcs}
\end{figure}

In order to illustrate this point, I will consider the
strictly non-relativistic reduction of the electromagnetic one-body
current operator. In section \ref{secres}, I also include the full
relativistic, positive-energy form as discussed in \cite{relcur}, but
for the moment, the familiar non-relativistic form is completely
sufficient to illustrate why and where factorization fails.  The
non-relativistic current operator consists of a charge part and of a
magnetization current and a convection current:
\begin{eqnarray}
J^o_{nonrel} &=& G_E \nonumber \\
 J^{\perp}_{nonrel} &=& - \frac{i}{2 \, m_N} \,  G_M \, \left (
\vec q \times \vec \sigma \right ) + \frac{1}{m_N} \, G_E \,
\left ( \vec p - \frac{\vec q \cdot \vec p}{q^2} \, \vec q
\right ) \,.
\end{eqnarray}
It is clear from the structure of the current operator that matrix
elements which contain the charge operator, $G_E$, or the
magnetization current, $- \frac{i}{2 \, m_N} \, G_M \, \left ( \vec q
\times \vec \sigma \right )$, differ only in the spin-structure, but
not in their structure in coordinate or momentum space.  So, as long
as the final state interaction operator is purely central,
factorization is valid for the matrix elements of the charge operator
and the magnetization current. However, the convection current
contains a gradient operator in coordinate space, coming from the
$\vec p_{\perp}$ in momentum space, and therefore, the matrix element
of the convection current differs from the other matrix elements in
coordinate or momentum space - it does not factorize. Now, the
validity of factorization depends on the importance of the convection
current contribution to the observable in question.  As the key
observable is the cross section, I will discuss the responses that
contribute to it. The convection current obviously does not contribute
to $R_L$, and the magnetization current is dominant in $R_T$, so one
might expect a factorization breaking of only a few percent in
$R_T$. So far, factorization would be acceptable, but there are also
the interference responses $R_{TL}$ and $R_{TT}$ which contribute to
the cross section.  While the interference responses are at least an
order of magnitude smaller for low missing momenta, they become
comparable to $R_L$ and $R_T$ for higher missing momenta, and are
therefore quite important for the cross section. From the spin
structure, it is clear that $R_{TL}$ in the non-relativistic approach
is proportional to the product of the charge operator and convection
current matrix elements. Due to the presence of the convection current,
 it is not going to factorize. The same holds for $R_{TT}$, as this
response contains only the convection current matrix elements. So in
the general case, even for the simple non-relativistic current
operator, factorization will not hold for higher missing momenta.
Factorization will be approximately valid in parallel and
anti-parallel kinematics, as the interference responses do not
contribute there. In the next section, I will show results for a
relativistic current operator. There, it will be obvious that
factorization works even less well in the relativistic case.
The relativistic operator
contains additional, non-factorizing operator structures and new
coefficients for the old operator structures, which may contain
kinematic factors like $\vec p \, ^2$, which break factorization, too.
For now, I will just show the results for the validity of
factorization in the non-relativistic approach for several different
kinematic settings. In Fig.~\ref{figratnrcs}, I show the ratio of the
cross section calculated in the factorization approximation to the
unfactorized cross section in parallel and perpendicular kinematics
for fixed kinetic energy of 1 GeV for the outgoing proton, and for
different values of the energy transfer $\omega$ for fixed
three-momentum transfer $|\vec q| = 2.0$ GeV/$c$.

From the left panel in Fig.~\ref{figratnrcs}, it is clear that the
violation of factorization is smallest in (anti)parallel kinematics.
There, the interference responses vanish and the breaking of
factorization stems solely from the small convection current
contribution to the transverse response. The resulting deviation from
1 of the ratio is of the order of a few percent only.  In
perpendicular kinematics, the deviations from 1 are much larger.  The
two curves shown for $\theta_m = 90^o$ differ by the azimuthal angle
$\phi_m$ of the neutron. The ejected nucleon's azimuthal angle is
$\phi = \phi_m + \pi$. The response $R_{TL}$ implicitly contains a
$\cos (\phi)$ dependence.  So, the only difference in the two cross
sections is that in one case, the transverse-longitudinal interference
response is added, and in the other case, it is subtracted from the
sum of the other responses. The breaking of factorization increases
with the missing momentum. This can be understood as FSI is mainly
responsible for the factorization breaking. At the energies
considered here, FSI is mainly diffractive and short-ranged, so that
it leads to large contributions at large missing momenta (see
e.g. \cite{myoldpapers,sofsi}). Also, the interference responses become
comparable to the other responses at larger $p_m$.  In perpendicular
kinematics, the deviations from 1 are considerable for $p_m > 300$
MeV/$c$ and range from 5 \% to 15 \%.  A comparison of perpendicular
kinematics and (anti)parallel kinematics clearly shows that the
contribution of the interference responses leads to strong
factorization breaking.

In the right panel of Fig.~\ref{figratnrcs}, the situation is
depicted for fixed energy and three-momentum transfer.  In such a
setting, the angle of the missing momentum changes with $p_m$.
Therefore, the interference responses are present in the cross
section, and the factorization breaking is noticeable for $p_m > 400$
MeV/$c$. For larger missing momenta, deviations range roughly from 6
\% to 9 \% and seem to grow with increasing transferred energy.  This
comes about as $R_{TL}$ increases with the energy transfer.

After identifying the source of factorization breaking due to FSI and
discussing the mechanism for the (too) simple case of a strictly
non-relativistic current operator, I proceed to give a realistic
estimate of the validity of the factorization approximation in the
next section.

\section{Realistic numerical examples}
\label{secres}

In this section, I use the relativistic, on-shell form, positive
energy (OSPE) current operator discussed in \cite{relcur,origins}.
Using this form of the current operator, I choose a specific off-shell
prescription. Currently, there exists no microscopic description of
the off-shell behavior that can be applied for a wide range of
kinematic conditions - there are only ad hoc prescriptions
\cite{deforest,koch,cdp}.  Here, I use the popular ansatz of applying
the electromagnetic current operator in its on-shell form. In
principle, one can perform the same analysis of the validity of
factorization using a current operator with a more general off-shell
behavior. However, there is no reason to assume that factorization
would work any better with a more general -- and therefore more
complicated -- off-shell behavior.  The OSPE current has the following
form:
\begin{equation}
J^\mu (P\Lambda ;P'\Lambda ^{\prime }) \equiv 
\chi _{\Lambda
^{\prime }}^{\dagger }  \, \, \bar{J}^{\mu }(P;P^{\prime }) 
\, \, \chi _{\Lambda }^{{}}
\end{equation}
with
\begin{eqnarray}
\bar J^o &=& \rho = f_o \left (
\xi _o + \, i \, \, \xi _o^{\prime } \, \left( \vec q \times \vec p
\right) \cdot \vec{\sigma }   \right ) \nonumber \\
\bar J^3 &=& \, \,  \frac{\omega}{q} \, \, \bar J^o  \nonumber \\
\bar J^{\bot } &=& f_o \left (  \xi _{1} \left[ \, \vec p
-\left( \frac{\vec q
\cdot \vec p}{q ^2}\right) \vec q \, \right] -i
\left\{ \xi _1^{\prime } \left( \vec q \times \vec{\sigma }
\right) \right.  \right. \nonumber\\
  &+&  \left. \left. \xi _2^{\prime } \left( \vec q \cdot \vec{\sigma }
\right) \left( \vec q \times \vec p \right) +\xi
_3^{\prime }\left[ \left( \vec q \times \vec p \right)
\cdot \vec{\sigma }\right] \left[ \vec p -\left( \frac{\vec q 
\cdot \vec p }{q ^2}\right) \vec q \right]
\right\} \right )  .
\label{jrel}
\end{eqnarray}
Here, $f_o, \xi_i, \xi_i'$ are all functions of $\omega, q, p^2$;
their explicit forms are:
\begin{eqnarray}
f_{0} &\equiv &\frac{1}{\mu _{1}\sqrt{1+\frac{\tau }{4\left( 1+\tau \right) }%
\mu _{2}^{2}\delta ^{2}}},  \label{sn13}
\end{eqnarray}
\begin{eqnarray}
\xi _{0} &=&\frac{\kappa }{\sqrt{\tau }}\left[ G_{E}+\frac{\mu _{1}\mu _{2}}{%
2(1+\tau )}\delta ^{2}\tau G_{M}\right]   \nonumber \\
\xi _{0}^{\prime } &=&\frac{1}{\sqrt{1+\tau }}\left[ \mu _{1}G_{M}-\frac{1}{2%
}\mu _{2}G_{E}\right]   \nonumber \\
\xi _{1} &=&\frac{1}{\sqrt{1+\tau }}\left[ \mu _{1}G_{E}+\frac{1}{2}\mu
_{2}\tau G_{M}\right]   \nonumber \\
\xi _{1}^{\prime } &=&\frac{\sqrt{\tau }}{\kappa }\left( 1-\frac{\mu _{1}\mu
_{2}}{2(1+\tau )}\delta ^{2}\right) G_{M}  \nonumber \\
\xi _{2}^{\prime } &=& \frac{\lambda \sqrt{\tau }}{2\kappa ^{3}}\mu _{1}\mu
_{2}G_{M}  \nonumber \\
\xi _{3}^{\prime } &=&\frac{\sqrt{\tau }}{2\kappa (1+\tau )}\mu _{1}\mu
_{2}\left[ G_{E}-G_{M}\right] . 
 \label{sn17}
\end{eqnarray}
The dimensionless variables are defined as follows:
\begin{eqnarray}
\kappa &=& \frac{|\vec q|}{ 2 m_N} \nonumber \\ 
\delta &=& \frac{p_{\perp}}{m_N} \nonumber \\ 
\tau &=& \kappa ^2 - \lambda ^2 \nonumber \\ 
\lambda &=& \frac{\omega}{2 m_N} \,
\end{eqnarray}
and $\mu_1, \mu_2$ are shorthand for
\begin{eqnarray}
\mu _1 &\equiv &\frac{\kappa \sqrt{1+\tau }}{\sqrt{\tau }\left( \varepsilon
+\lambda \right) }=\frac 1{\sqrt{1+\frac{\delta ^2}{1+\tau }}}  \label{sn8}
\\
\mu _2 &\equiv &\frac{2\kappa \sqrt{1+\tau }}{\sqrt{\tau }\left( 1+\tau
+\varepsilon +\lambda \right) }=\frac{2\mu _1}{1+\frac{\sqrt{\tau (1+\tau )}}%
\kappa \mu _1}. 
\end{eqnarray}

For the reasons explained in \cite{relcur}, I refer to the operator
associated with $\xi_o$ as zeroth-order charge operator, I call the
term containing the $\xi_o'$ first-order spin-orbit operator, the term
containing $\xi_1$ first-order convection current, the term containing
$\xi_1'$ zeroth-order magnetization current, the term containing
$\xi_2'$ first-order convective spin-orbit term, and the term
containing $\xi_3'$ second-order convective spin-orbit term.  In this
paper, the current is used in this unexpanded, full form, which is
possible as the evaluation of the FSI integrals takes place in
momentum space.  Some technical problems pertaining to the
coordinate space treatment can be avoided this way.

\begin{figure}[!h,t]
\begin{center}
\leavevmode
\epsfig{file = 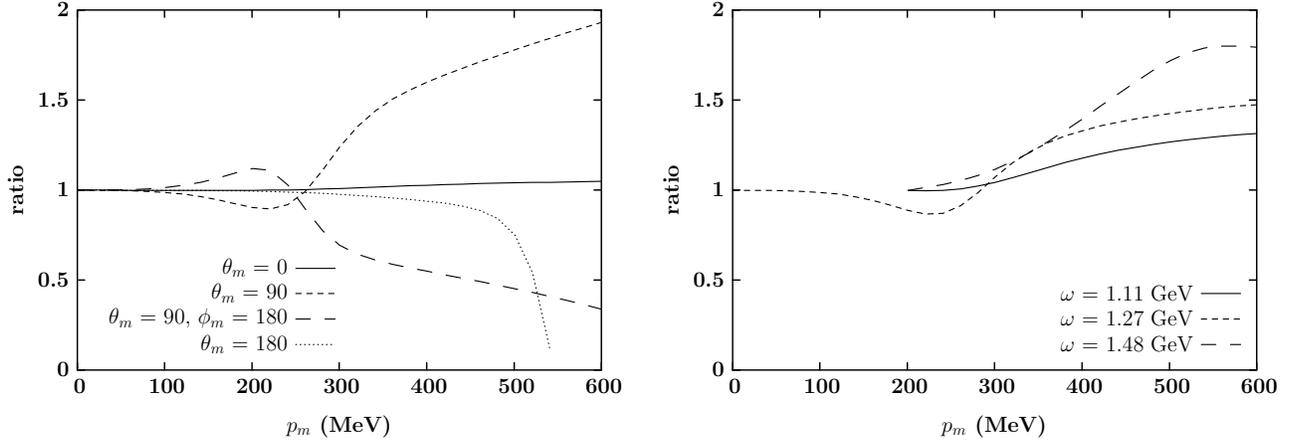, height = 16cm}
\end{center}
\vspace{-9cm}
\caption{The ratio of the cross section calculated in the
factorization approximation to the unfactorized cross section for
different kinematic settings.  The full relativistic form of the
current operator was employed.  The left panel shows the ratio for
fixed kinetic energy of 1 GeV of the outgoing proton and various fixed
angles of the missing momentum, the right panel shows the ratio for
fixed 3-momentum transfer $q = 2.0$ GeV/c and different values of the
fixed energy transfers $\omega$: 1.11 GeV, 1.27 GeV, and 1.48 GeV.
Note that the scales in this figure are different from the ones in
Fig.~\protect{\ref{figratnrcs}}.}
\label{figratexcs}
\end{figure}

The final state interaction is calculated using Glauber theory, see
e.g. \cite{sofsi}. For the purpose of this paper, considering the
central, dominating part of the FSI is sufficient, as the breaking of
factorization is strong already in this case. Spin-dependent FSI will
break factorization even for the non-relativistic forms of the charge
operator and the magnetization current. However, it is quite small
compared to the central FSI, and makes its major contribution to the
smallest response, $R_{TT}$, and to the fifth response, which does not
enter the unpolarized quantities I consider here. In other words, the
case against factorization is obvious already from using only central
FSI, and any spin-dependent FSI will only increase the problem.  In
this paper, I use only the central FSI for simplicity, although
calculations including the spin-orbit FSI are available in the
literature, see e.g. \cite{miller,iks,sofsi}. Glauber theory is the
main tool used for calculating FSI at GeV energies. The details of the
employed FSI operators and the parameters used for it are not
important for the current purpose, as the breaking of factorization
depends only on the presence of final state interaction. The detailed
form of the current operator is much more important, as can be seen
from the comparison of Fig.~\ref{figratnrcs} and
Fig.~\ref{figratexcs}.

In Fig.~\ref{figratexcs}, I show the ratio of the factorized to the
unfactorized cross section for the same kinematic conditions as in
Fig.~\ref{figratnrcs}, but with the full relativistic OSPE current of
Eq. (\ref{jrel}). Note that the scales are different in the two
figures in order to accommodate the larger deviations from unity in the
relativistic case.  Comparing the two figures, it is obvious that the
more complicated structure of the relativistic current operator leads
to a much larger breaking of the factorization assumption in all
considered kinematics.

The ratio in parallel and anti-parallel kinematics (left panel) is
still rather close to 1, deviations at higher missing momentum are of
the order of 5\%. The larger deviations in anti-parallel kinematics
occur close to the kinematic threshold (values larger than a certain
$p_{m,max}$ cannot be reached for a fixed proton momentum, which is
implied by a fixed proton kinetic energy), and are not of great
practical relevance. The only responses contributing in these
kinematics are $R_L$ and $R_T$, and the deviations from 1 now stem not
only from the convection current, but also from the first and
second-order convective spin-orbit contributions to the transverse
part of the current operator and from the spin-orbit operator in the
charge operator. In addition, the factors $\xi_0$ and $\xi_1'$ which
multiply the zeroth-order charge operator and the magnetization
current depend on $\delta^2 = \frac{p_{\perp}^2}{m^2}$, and therefore
not even the matrix elements $<f|\hat S_{FSI} J_{em}|i>$ of the
zeroth-order charge operator and the magnetization current are
proportional, and factorization does not hold at all in this
relativistic setting. This is reflected by the larger amount of
factorization breaking in the relativistic case,
Fig.~\ref{figratexcs}, compared to the non-relativistic case,
Fig.~\ref{figratnrcs}. Factorization breaking on the order of 5\%, as
observed in parallel kinematics, is not a terribly large
effect. 
However, one needs to take into account that in an actual experiment,
exactly parallel kinematics are not necessarily achieved, and that
sometimes data corresponding to a larger range in acceptance may be
combined in a single bin. E.g. for $\theta_m$ = 10$^o$, the deviations are
rising to 6 \% at missing momenta around 400 MeV/$c$ and to 11 \% at
missing momenta around 600 MeV/$c$. For $\theta_m$ = 15$^o$, the
deviations are rising to 8 \% at missing momenta around 400 MeV/$c$
and to 14 \% at missing momenta around 600 MeV/$c$.
%
%
While this is still good enough for count rate estimates, one
certainly does not want to incur this error in a precise theoretical
prediction by making an entirely unnecessary approximation like
factorization.

In perpendicular kinematics (dashed curves, left panel), the
deviations from 1 are now large for missing momenta $p_m >$ 300
MeV/$c$, and they are non-negligible for missing momenta from 100
MeV/$c$ to 300 MeV/$c$.  In addition to the factorization breaking in
$R_L$ and $R_T$, the interference responses contribute strongly to the
factorization breaking. The reason for the huge increase in
factorization breaking going from the non-relativistic to the
relativistic treatment is that in the relativistic treatment, the
interference responses pick up large contributions \cite{relcur}, and
are now much more important in the cross section, specifically for
large missing momenta. 

When fixing transferred energy and three-momentum (right panel), the
factorization breaking is present for missing momenta from 200 to 300
MeV/$c$, and very large for missing momenta beyond that. In these
kinematics, the missing momentum angle varies, so that at any value of
$p_m$, one can expect the interference responses to
contribute. Therefore, the factorization breaking is large, even
though not quite as large as in perpendicular kinematics, where the
contribution of the interference responses is maximized.  Again, one
sees a large increase in factorization breaking going from the
non-relativistic case to the relativistic case, due to the
interference responses.

\section{Summary and Conclusions}

I have pointed out the sources of factorization breaking in $(e,e'p)$
reactions at GeV energies, and given numerical examples for the
reaction $^2H(e,e'p)$. Both in the (oversimplifying) non-relativistic
treatment and in the relativistic case, the factorization breaking is
significant. The strength of the factorization breaking depends
considerably on the chosen kinematics.  Only at very low missing
momenta, $p_m <$ 100 MeV/$c$, factorization works. In strictly
parallel kinematics, the deviations are about 5\% or smaller. However,
one needs to keep in mind that in an experiment, a range of angles
around $0^o$ may contribute, and the deviations will be
correspondingly larger, around 10 \% for large missing momenta.  In
perpendicular kinematics or for fixed transferred energy and
three-momentum, the factorization assumption clearly fails for $p_m >$
300 MeV/$c$.

While it is well known that factorization is insufficient when
$(e,e'p)$ reactions at lower energies are calculated, this fact does
not seem to be widely appreciated when it comes to GeV energy and
momentum scales. This paper serves to draw attention to the fact that
this approximation is lacking and that correct treatments of at least
this problem are available, both in the GeV regime
and in the transition region from lower to higher energies,
see e.g. \cite{janrel,miller,iks,sofsi,udias,wallynf}. It is
especially important to be accurate as far as factorization is
concerned, as there are other aspects of the problem,
e.g. relativistic two-body currents, which are not yet worked out to a
satisfactory degree, and which are going to cause uncertainties in the
theoretical calculations. Note that many color transparency calculations,
see e.g. \cite{kolya,fs}, assume factorization. The color
transparency effects predicted for experiments at Jefferson Lab are
relatively small, and the additional uncertainty introduced by assuming 
factorization may very well be of the same order of magnitude
as the predicted effects and rather misleading in the
interpretation of the data.

In this paper, I have considered the effects of factorization on the
unpolarized cross section only. It is clear that e.g. polarization
observables or single responses are more sensitive to this type of
approximation.  One hardly needs to point out that wherever a dip is
predicted for an observable in a factorized calculation, it will most
likely be filled in when the correct, unfactorized calculation is
performed. Furthermore, interesting new information will be obtained
from separating the responses $R_L + R_{TT}$, $R_T$, and $R_{TL}$,
which cannot be interpreted in a factorized approach.

\acknowledgments

I would like to thank T. W. Donnelly, N. N. Nikolaev, and J. W. Van Orden
for many stimulating discussions on the subject of electron
scattering.  I am grateful to Shalev Gilad for clarifying
some points concerning the experimental set-up which I had misunderstood.
I would like to thank R. Schiavilla for providing a
parameterization of the Argonne V18 deuteron wave functions.  This
work was supported by funds provided by the U.S. Department of Energy
(D.O.E.)  under cooperative research agreement \#DE-AC05-84ER40150.

\end{document}